\documentclass{article}
\usepackage{a4wide}
\usepackage{amssymb}
\usepackage{graphicx}
\usepackage{morefloats} 
\usepackage{color}
\usepackage{amsmath}
\usepackage{amsfonts}
\usepackage{amssymb}
\usepackage{hyperref}
\usepackage[colorinlistoftodos]{todonotes}
\usepackage{algorithm}
\usepackage{algorithmic}
\usepackage{fix-cm}
\usepackage[normalem]{ulem}

\def\bfx{\mbox{\boldmath$x$}}
\def\bfv{\mbox{\boldmath$v$}}

\usepackage{booktabs}

\begin{document}

%\begin{frontmatter}



%\dochead{International Conference on Computational Science, ICCS 2013}
%% e.g. \dochead{17th International Conference on Dynamical Processes in Excited States of Solids}

\title{Initialization of lattice Boltzmann models with the help of the numerical Chapman--Enskog expansion}


\author{Y.~Vanderhoydonc and W.~Vanroose \\ \small{Dept.~Mathematics and Computer Science, Universiteit Antwerpen}}
\date{}

\maketitle

\begin{abstract}
We extend the applicability of the numerical Chapman--Enskog expansion as a lifting operator for lattice Boltzmann models to map density and momentum to distribution functions. In earlier work [Vanderhoydonc et al.~Multiscale Model.~Simul.~10(3): 766-791, 2012] such an expansion was constructed in the context of lifting only the zeroth order velocity moment, namely the density. A lifting operator is necessary to convert information from the macroscopic to the mesoscopic scale. This operator is used for the initialization of lattice Boltzmann models. Given only density and momentum, the goal is to initialize the distribution functions of lattice Boltzmann models. For this initialization, the numerical Chapman--Enskog expansion is used in this paper.
 \\\\ \small{ \it{Keywords: Numerical Chapman--Enskog expansion; lifting operator; lattice Boltzmann models; Chapman--Enskog expansion; Constrained Runs; macroscopic partial differential equations; macroscopic velocity moments. }}
 \end{abstract}




% \end{keyword}
%  \cortext[cor1]{Corresponding author. Tel.: +3232653859 ; fax: +3232653777.}
% \end{frontmatter}

%\correspondingauthor[*]{Corresponding author. Tel.: +0-000-000-0000 ; fax: +0-000-000-0000 .}
%\email{ynte.vanderhoydonc@ua.ac.be}

% \linenumbers

\section{Introduction}
The dynamics of a system of colliding particles can typically be described by different levels of accuracy. One distinguishes between micro-, meso- and macroscopic scales. Macroscopic partial differential equations (PDEs) model only a few low order velocity moments and are therefore not that accurate to describe interactions between particles. When a detailed description is necessary, a microscopic model is used. These individual-based models can take into account the particle collision physics. Kinetic equations are ubiquitous to model at this scale. The Boltzmann equation can be used to describe such kinetic models by modelling distribution functions in phase space. 

We will focus on lattice Boltzmann models (LBMs) that are space, time and velocity discretizations of the Boltzmann equation \cite{LGCA_LBM, succi}. LBMs are used for hydrodynamic simulations of complex fluids \cite{LGCA_LBM, succi, aidun} and are based on a streaming and collision process. The initialization of such an LBM is performed by using a lifting operator. Such an operator defines a mapping between macroscopic and mesoscopic variables such that the distribution functions of the LBM can be built from the given macroscopic information. The concept of a lifting operator in a multiscale context was first introduced by Kevrekidis et al.~in the equation-free framework to couple different scales in a dynamical system \cite{kevrekidis}.

Lifting operators for LBMs are considered in \cite{leemput_phd, leemput, vanderhoydonc1, vanderhoydonc2, initialization, vandekerckhove} where density is mapped to distribution functions. The Chapman--Enskog expansion is built for such model problems in \cite{leemput_phd}. This expansion is first introduced in \cite{chapman_cowling} to solve the Boltzmann equation. Furthermore, \cite{leemput_phd}, \cite{leemput}, \cite{initialization} and \cite{vandekerckhove} apply the Constrained Runs (CR) algorithm to LBMs. Originally, the CR algorithm is introduced by Gear et al.~\cite{gear} to map macroscopic initial variables to missing microscopic variables for stiff singularly perturbed ordinary differential equations (ODEs). This algorithm is based on the attraction of the dynamics toward the slow manifold. Since the dynamics on this slow manifold can be parametrized by only macroscopic variables such as the density, the higher order velocity moments become slaved functionals of the density in the LBM context with conserved density. \cite{vanderhoydonc1} compares the Chapman--Enskog expansion with the CR algorithm for model problems with given initial density. However, both of these methods have some serious drawbacks. The Chapman--Enskog expansion needs to be constructed analytically while the CR algorithm is computationally expensive, especially for higher dimensional problems. Because of these drawbacks, \cite{vanderhoydonc2} constructs the numerical Chapman--Enskog expansion as a lifting operator for these model problems. This numerical Chapman--Enskog expansion is based on a combination of the Chapman--Enskog expansion and the CR algorithm. It finds the coefficients of the Chapman--Enskog expansion numerically based on the CR algorithm. This reduces the number of unknowns in the lifting since it only finds the coefficients of the expansion rather than the full state of the distribution functions.

So far, the numerical Chapman--Enskog expansion has been constructed for model problems with equilibrium distribution functions that only depend on the density. They appear in LBMs where the collisions only conserve the density. The goal of this paper is to generalize it to models where both density and momentum are conserved.
We borrow the concept of the numerical Chapman--Enskog expansion constructed in \cite{vanderhoydonc2} and use it to map both density and momentum to distribution functions of the LBM. The proposed algorithm is an extension that shows that the numerical Chapman--Enskog expansion can be generalized for the considered model problems. These model problems still face limiting assumptions but are an intermediate step towards more general physical problems.

The remainder of this paper is organized as follows. Section \ref{diff_models} starts with the outline of the models with different levels of accuracy in which we are interested. Both lattice Boltzmann models and more macroscopic PDEs are briefly discussed.
Section \ref{NCHE_density} describes the numerical Chapman--Enskog expansion to map density to distribution functions.
We then proceed to the construction of the numerical Chapman--Enskog expansion to map both density and momentum to distribution functions in Section \ref{NCHE_density_momentum}.
Numerical results for academic test problems are considered in Section \ref{numerical_results}. There the proposed lifting operator is tested in a setting of restriction and lifting.
We conclude and give an outlook in Section \ref{conclusions}.

\section{Models with different levels of accuracy} 
\label{diff_models} 
This section introduces models that will be used throughout this
paper. The mesoscopic scale is described by a lattice Boltzmann model (LBM)
and will be discussed in Section \ref{LBMs}. The macroscopic scale
will use partial differential equations (PDEs) to describe the
evolution of the macroscopic velocity moments in Section
\ref{macro_eq}.

\subsection{ Lattice Boltzmann models } \label{LBMs}
An LBM \cite{LGCA_LBM, succi} describes the
evolution of a one-particle distribution function
$f_i(\bfx,t)=f(\bfx,\bfv_i,t)$ discretized in space $\bfx \in
\mathbb{R}^n$, time $t \geq 0$ and velocity $\bfv_i \in \mathbb{R}^n$
with $n \in \{1,2,3\}$ for practical applications. The velocities are
taken from a discrete set defined by the geometry of the grid. The
functions are represented as $f_i: \mathcal{X} \times \mathcal{T}
\rightarrow \mathbb{R}$ with $\mathcal{X} \times \mathcal{T} $ the
space-time grid with space steps $\Delta \bfx_i$ in the direction of
velocity $\bfv_i$, time step $\Delta t$ and $\mathcal{T}=\{0,\Delta t,
2\Delta t,\ldots\} \cap [0,T]$.  Representation DdQq used for the
description of LBMs stands for d dimensions and q velocity directions.
D1Q3, for example, considers a one-dimensional spatial domain with
only three values for the velocity $v_i=c_i \Delta x/\Delta t$ in
which $c_i=i$, $i \in \{-1,0,1\}$ represents the dimensionless grid
velocities.  The remainder of this section contains the description of
the D1Q3 lattice Boltzmann equation (LBE). This
equation can however easily be generalized to more dimensions.

The lattice Boltzmann equation describing the evolution of the
distribution functions 
with BGK approximation \cite{BGK} and no external forces is
\begin{equation}\label{LBE}
f_i(x+c_i\Delta x,t+\Delta t)=(1-\omega)f_i(x,t)+\omega f_i^{eq}(x,t), \quad i \in
\{-1,0,1\}.
\end{equation}
The equilibrium distribution functions $f_i^{eq}(x,t)$ are defined to satisfy the conservation laws. For
example, when the LBM only conserves the density $\rho(x,t)$, equilibrium
distribution functions can be defined by
$f_i^{eq}(x,t)=\frac{1}{3}\rho(x,t)$. The relaxation time
$\tau=1/\omega$ represents the time scale associated with the
relaxation towards equilibrium.

\subsection{Macroscopic partial differential equations} \label{macro_eq}

At the macroscopic scale the evolution is described by a PDE for the
moments of the particle distribution functions.  Density
$\rho(x,t)=\sum_i f_i(x,t)$, momentum $\phi(x,t)=\rho(x,t)u(x,t)=\sum_i
v_if_i(x,t)$ and energy $\xi(x,t)=\frac{1}{2}\sum_i v_i^2 f_i(x,t)$
are respectively the zeroth, first and second order velocity moments.

For the D1Q3 problem, discussed in this section, the transformation
between the distribution functions and the moments is straightforward
since the matrix $\boldsymbol{M}$ below is invertible.
\begin{equation}\label{omzet}
\left(\begin{array}{c}
\rho \\
\phi \\
\xi
\end{array}\right)
=
\left(\begin{array}{c c c}
1 & 1 & 1 \\
1 & 0 & -1 \\
\frac{1}{2} & 0 & \frac{1}{2}
\end{array}\right)
\left(\begin{array}{c}
f_{1} \\
f_0 \\
f_{-1}
\end{array}\right)
=\boldsymbol{M}
\left(\begin{array}{c}
f_{1} \\
f_0 \\
f_{-1}
\end{array}\right).
\end{equation}
When we look at these functions in a point $x$ at time $t$, they can
be represented either as $(f_{1}, f_0, f_{-1})^T \in \mathbb{R}^3$ or
as $(\rho, \phi, \xi)^T \in \mathbb{R}^3$. 

Low order velocity moments are used to describe macroscopic
PDEs. For example, it can be shown that the diffusion PDE and the LBM
are macroscopic equivalent \cite{leemput_phd} when considering the following.
\begin{equation}\label{link_PDE_LBM}
\frac{\partial \rho}{\partial t}=D\frac{\partial^2\rho}{\partial x^2}, \quad D=\frac{2-\omega}{3\omega}\frac{\Delta x^2}{\Delta t}, \quad f_i^{eq}(x,t)=\frac{1}{3}\rho(x,t), \quad i \in \{-1,0,1\}.
\end{equation}

\begin{figure}[h]
\includegraphics[width=\textwidth]{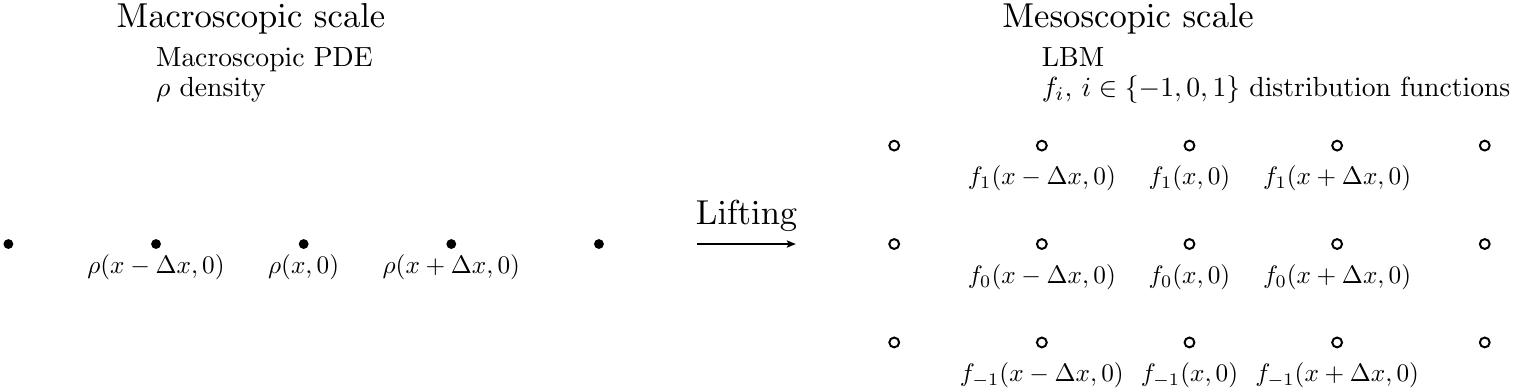}
\caption{To initialize an LBM from a given density $\rho(x,0)$, or density and momentum, we need to characterize the distribution functions $f_i(x,0)$, $i \in \{-1,0,1\}$ at time $t=0$. A lifting operator is then used to map the macroscopic variables to the mesoscopic variables, namely the distribution functions of the LBM. \label{fig.scales}}
\end{figure}

Figure \ref{fig.scales} clarifies the intention of this paper, namely the initialization of LBMs. From a given density $\rho(x,0)$, or density and momentum, determine the distribution functions $f_i(x,0)$, $i \in \{-1,0,1\}$ of the LBM at time $t=0$. This will include some arbitrariness. To fully characterize
the distribution functions $f_i(x,0)$ at time $t=0$ all three moments --- 
$\rho(x,0)$, $\phi(x,0)$ and $\xi(x,0)$ --- are necessary. 
For this, a lifting operator is necessary that defines a mapping between macroscopic and mesoscopic variables such that the distribution
functions of the LBM can be built from the given macroscopic information.

The aim is then
to find lifting methods that predict the missing moments so that steep
initial layers that occur in ill-initialized problems are avoided
\cite{mei}. The focus of this paper is to initialize LBMs from
a given density and momentum, the zeroth and first order velocity
moments, using a numerical Chapman--Enskog expansion.  Such an  expansion
combines the ideas of the Chapman--Enskog expansion
\cite{chapman_cowling} and the Constrained Runs (CR) algorithm
\cite{gear}. Both methods are used in the literature as lifting
operators for LBMs. The Chapman--Enskog expansion for LBMs is outlined
in \cite{leemput_phd} and the application of CR
to LBMs is considered in \cite{initialization}. The numerical Chapman--Enskog expansion was first
proposed in \cite{vanderhoydonc2} for LBMs that only conserve the
density.

% Symbols denoting vectors and matrices should be indicated in bold type. Scalar variable names should normally be expressed using italics.

\section{Numerical Chapman--Enskog expansion for LBMs to map density to
  distribution functions} \label{NCHE_density} 

This section describes the numerical Chapman--Enskog expansion for problems
where the equilibrium distribution functions only depend on the
density. The goal of this paper is to generalize this to a dependence
of both density and momentum. The description is based on D1Q3 problems but
this can easily be generalized to higher dimensional problems. 

\subsection{Chapman--Enskog expansion}
The Chapman--Enskog expansion expands distribution functions around the equilibrium as
\begin{equation*}
f=f^{(0)}+\varepsilon f^{(1)}+\varepsilon^2 f^{(2)}+\ldots,
\end{equation*}
where $f^{(1)}$ and $f^{(2)}$ are correction terms \cite{chapman_cowling}. For D1Q3 problems,
where the equilibrium only depends on the density, the correction
terms form a series of the density and its derivatives
\cite{leemput_phd, leemput}. Thus this expansion can be used to map the 
density to the distribution functions. 

However, the derivation of the correction terms requires tedious
analytical work, especially higher order terms for multidimensional
problems are not straightforward to obtain.

\subsection{Constrained Runs algorithm}
The CR algorithm is a numerical alternative based on the
attraction of the dynamics toward the slow manifold. It was first
introduced by Gear et al.~to map macroscopic initial variables to
missing microscopic variables for stiff singularly perturbed ordinary
differential equations (ODEs) \cite{gear}.  Applied to
LBMs, the CR algorithm initializes the missing velocity moments such that the
evolution of these moments is smooth of order $m$. The smoothness
condition is defined by
\begin{equation} \label{CR_formulas}
\frac{d^{m+1}\phi(t)}{dt^{m+1}}\bigg|_{t=0}=0 \quad \text{and} \quad \frac{d^{m+1}\xi(t)}{dt^{m+1}}\bigg|_{t=0}=0.
\end{equation}
The higher the order of smoothness, the more accurate the method. 

In practice the smoothness condition is approximated by a finite
difference relation between the moments in successive time steps.
There are two possible ways to find the moments that satisfy the
smoothness condition. Via an explicit iteration but this can become
unstable for the higher order smoothness condition \cite{leemput_phd, vandekerckhove}.  Or via a
Newton iteration that solves the fixed point problem
\begin{equation*}
\left(\begin{array}{c}
\phi \\
\xi
\end{array}\right)^{k+1} = \mathcal{C}_m \left(\rho_0,\left(\begin{array}{c}
\phi \\
\xi
\end{array} \right)^{k}\right),
\end{equation*}
where $\mathcal{C}_m$ is a backward extrapolation formula derived from
the smoothness condition and $\rho_0$ is the given density \cite{vandekerckhove}.  More stability and
accuracy details concerning the application of CR to
LBMs can be found in \cite{leemput_phd, initialization, vandekerckhove}. 

This CR algorithm has a serious computational cost since it
searches for the missing moments in every grid point of the spatial domain.  Especially for high dimensional
problems this leads to large Jacobian matrices that need to be
estimated through additional LBM steps. This makes the CR algorithm
prohibitively expensive for 2D problems.

\subsection{Numerical Chapman--Enskog expansion}
Instead of using CR to find for each grid point the
missing moments $\phi$ and $\xi$, the numerical Chapman--Enskog
expansion uses CR to find the unknown coefficients of
the expansion \cite{vanderhoydonc2}. The major advantage is that it
leads to much smaller systems since it finds the expansion
coefficients that satisfy the smoothness condition rather than the
full state of the velocity moments. This reduces the number of
unknowns significantly. The numerical Chapman--Enskog expansion is
described in this section for one-dimensional problems, with
equilibrium distribution functions that only depend on the density. This
can also easily be generalized to more dimensions.

The expansion expresses the distribution functions as
\begin{equation} \label{constants}
  f_i(x,t)  =  f_i^{eq}(x,t) + \alpha_i \partial_x \rho + \beta_i  \partial^2_x \rho  + \ldots \nonumber
+ \gamma_i \partial_t \rho + \zeta_i \partial_t^2 \rho + \ldots + \eta_i \partial_x \partial_t \rho + \ldots,
\end{equation}
where \begin{equation} \label{vectors_constants}
i \in \{-1,0,1\} \quad \text{and} \quad
\boldsymbol{\alpha}=
\left(\begin{array}{c}
\alpha_1 \\
\alpha_0 \\
\alpha_{-1}
\end{array}\right) \in \mathbb{R}^3, \quad
\boldsymbol{\beta}=
\left(\begin{array}{c}
\beta_1 \\
\beta_0 \\
\beta_{-1}
\end{array}\right) \in \mathbb{R}^3,
 \ldots,
\end{equation}
are vectors of constants that only depend on $\omega$, $\Delta x$ and
$\Delta t$. This expansion is valid for an LBM on an infinite domain
with parameters $\Delta x$, $\Delta t$ and $\omega$ and equilibrium
distribution functions that only depend on $\rho(x,t)$. It requires
some smoothness condition for the expansion to be valid. The derivation of this
expansion and the necessary conditions are outlined in
\cite{vanderhoydonc2}.

Once the constants are determined, the lifting operator --- that is
necessary to initialize the LBM --- can be applied repeatedly for the full domain to lift all the missing moments at a
minimal cost of estimating the derivatives of the density with finite
difference approximations. How these constants are found is discussed below as a summary of
\cite{vanderhoydonc2}. 

To simplify the discussion and notation we
limit ourselves to a truncated series
\begin{equation} \label{eq_coeff}
f_i(x,t)=f_i^{eq}(x,t)+\alpha_i \partial_x \rho + \beta_i  \partial_x^2 \rho +\gamma_i \partial_t \rho, \quad i \in \{-1,0,1\}.
\end{equation}

If a PDE in closed form exists that describes the evolution of
  $\rho$ in the form of $\partial_t \rho + a \partial_x \rho = D \partial_x^2 \rho$, then it is possible to eliminate $\partial_t \rho$ from the expansion. The distribution functions are now series with only spatial derivatives. 
\begin{equation}\label{f_constant}
f_i(x,t)=f_i^{eq}(x,t)+\alpha_i\partial_x \rho+\beta_i\partial_x^2 \rho, \quad i \in \{-1,0,1\}.
\end{equation}
Bear in mind that it considers different coefficients for $\alpha_i$ and $\beta_i$ in Eq.~\eqref{eq_coeff} and Eq.~\eqref{f_constant}.

We can extract the coefficients $\alpha_i$ and $\beta_i$ from the linear system

\begin{eqnarray} \label{system_constants_spatial}
\left(\begin{array}{ccc|ccc}
\partial_x\rho(x_j) & & & \partial_x^2 \rho(x_j)&& \\
&\partial_x \rho(x_j) & & & \partial_x^2 \rho(x_j)& \\
&&\partial_x \rho(x_j) & & & \partial_x^2 \rho(x_j) \\
\hline
\partial_x \rho(x_k) & & & \partial_x^2 \rho(x_k)&&\\
&\partial_x \rho(x_k) & & & \partial_x^2 \rho(x_k)&\\
&&\partial_x \rho(x_k) & & & \partial_x^2 \rho(x_k) \\
\end{array}
\right)
\left(\begin{array}{c}
  \alpha_{1}\\
  \alpha_0\\
  \alpha_{-1}\\\hline
  \beta_{1}\\
 \beta_0\\
  \beta_{-1}\\
\end{array}
\right) \nonumber \\
=
\left(\begin{array}{c}
  f_{1}(x_j,t) - f^{eq}_1(x_j,t)\\
  f_0(x_j,t) - f^{eq}_0(x_j,t)\\
  f_{-1}(x_j,t) - f^{eq}_{-1}(x_j,t)\\\hline
  f_{1}(x_k,t) - f^{eq}_{1}(x_k,t)\\
  f_0(x_k,t) - f^{eq}_0(x_k,t)\\
  f_{-1}(x_k,t) - f^{eq}_{-1}(x_k,t)\\
\end{array}
\right),
\end{eqnarray}
once the distribution functions are close to the slow manifold since on the slow manifold the distribution functions are parametrized by the density \cite{vanderhoydonc2}. $x_j$ and $x_k$ represent grid points of the spatial domain.

To reach the slow manifold we combine CR with the extraction of the
coefficients.  Consider a numerical function $h(\boldsymbol{\alpha},\boldsymbol{\beta}; \rho,
m)$ as described in Function \ref{fun:h}.  This function takes as
input $\boldsymbol{\alpha}$ and $\boldsymbol{\beta}$, the vectors of coefficients of expansion
\eqref{f_constant}, and as parameters the fixed density $\rho$ and an
integer $m$, the order of the smoothness condition.  It first
constructs, with this input, a state $(f_1,f_0,f_{-1})^T$ with the help of series
\eqref{f_constant}. This state is then used as an initial state to
perform a sequence of multiple LBM steps.  For each of these steps we find the
corresponding moments $\phi$ and $\xi$ using Eq.~\eqref{omzet}.  With these moments and the
smoothness condition we find a new guess for $\phi$ and $\xi$ that is closer to the slow
manifold.  These new moments result in new vectors of coefficients $\boldsymbol{\alpha}$ and
$\boldsymbol{\beta}$, by applying the linear system in Eq.~\eqref{system_constants_spatial} on the distribution functions
$f_i$, $i \in \{-1,0,1\}$.

The idea is now to determine $\boldsymbol{\alpha}$ and $\boldsymbol{\beta}$ such that they are
invariant under this numerical function $h(\boldsymbol{\alpha},\boldsymbol{\beta} ;
\rho,m)$. Indeed, if the initial and final state can be described by
the same $\boldsymbol{\alpha}$ and $\boldsymbol{\beta}$ then the lifted distribution function is close to the slow
manifold since the evolution of the corresponding $\phi$ and $\xi$
satisfies the smoothness condition.

The vectors of coefficients $\boldsymbol{\alpha}$ and $\boldsymbol{\beta}$ that are invariant under the numerical function $h(\boldsymbol{\alpha},
\boldsymbol{\beta} ; \rho,m)$ are found through a Newton iteration \cite{vanderhoydonc2}.
 
%\begin{algorithm}[!ht]
\begin{algorithm}[h]
\floatname{algorithm}{Function}
  \caption{$h(\boldsymbol{\alpha},\boldsymbol{\beta} ; \rho, m)$ \label{fun:h}} 
  \label{algo:coefficients}
  \begin{algorithmic}[1]
  \REQUIRE Guess on vectors of coefficients $\boldsymbol{\alpha}$, $\boldsymbol{\beta}$, given density $\rho$, order $m$ to use in Eq.~\eqref{CR_formulas}.
    \STATE Construct lifting operator in Eq.~\eqref{f_constant}.
    \STATE Compute corresponding moments $\phi$ and $\xi$ by applying Eq.~\eqref{omzet}. 
	\STATE Perform $m+1$ LBM time steps to compute the finite difference approximations of Eq.~\eqref{CR_formulas}. This results in new moments $\phi$ and $\xi$ that are closer to the slow manifold.
\STATE Revert back to distribution functions $(f_1,f_0,f_{-1})^T$ by applying Eq.~\eqref{omzet}.
    \STATE Select grid points $x_j$ and $x_k$ to construct the linear system in Eq.~\eqref{system_constants_spatial}.
    \STATE Solve the system for $\boldsymbol{\alpha}$ and $\boldsymbol{\beta}$.
    \RETURN $\boldsymbol{\alpha}$, $\boldsymbol{\beta}$.
  \end{algorithmic}
\end{algorithm}

% \begin{figure}[h]
% \centerline{\includegraphics{fx1}\hspace*{5mm}\includegraphics{fx1}}
% \caption{(a) first picture; (b) second picture.}
% \end{figure}

\section{Numerical Chapman--Enskog expansion for LBMs to map density and momentum to distribution functions} \label{NCHE_density_momentum}

This section generalizes the idea of the numerical Chapman--Enskog
expansion of Section \ref{NCHE_density} where the LBM only conserves the
density and the dynamics are described by the density as the only macroscopic variable.  Now, we will look at LBMs where both density
and momentum are conserved and both need to be considered as
macroscopic variables.  An important question is how this affects the numerical
Chapman--Enskog expansion?

The notation in this section is also based on D1Q3 problems. Again,
this can easily be generalized to more dimensions.

\subsection{Numerical Chapman--Enskog expansion}

We recall the derivation of the numerical Chapman--Enskog expansion in \cite{vanderhoydonc2} to show that the generalized expansion is given by
\begin{equation} \label{NCHE_rho_rhou}
  f_i(x,t)  =  f_i^{eq}(x,t) + \alpha_i^{(1)} \partial_x \rho + \alpha_i^{(2)} \partial_x (\rho u) +  \beta_i^{(1)}  \partial_x^2 \rho  + \beta_i^{(2)} \partial_x^2 (\rho u) + \ldots + \gamma_i^{(1)} \partial_t \rho + \gamma_i^{(2)} \partial_t (\rho u) + \ldots,
\end{equation}

where \begin{equation} 
i\in \{-1,0,1\} \quad \text{and} \quad
\boldsymbol{\alpha}^{(1)}=
\left(\begin{array}{c}
\alpha_1 \\
\alpha_0 \\
\alpha_{-1}
\end{array}\right)^{(1)} \in \mathbb{R}^3, \quad
\boldsymbol{\alpha}^{(2)}=
\left(\begin{array}{c}
\alpha_1 \\
\alpha_0 \\
\alpha_{-1}
\end{array}\right)^{(2)} \in \mathbb{R}^3,
 \ldots,
\end{equation}
are vectors of constants that only depend on $\omega$, $\Delta x$ and
$\Delta t$. 

Indeed, if the functions $f_i(x,t)$ are infinitely differentiable, a Taylor
expansion can be constructed and the distribution functions in point
$x+i\Delta x$, $i \in \{-1,0,1\}$ at time $t+\Delta t$ are given by
\begin{eqnarray}
f_i(x+i\Delta x, t+\Delta t)=f_i(x,t)+\frac{\partial f_i(x,t)}{\partial x}
i\Delta x+\frac{\partial^2 f_i(x,t)}{\partial x^2} \frac{i^2 \Delta x^2}{2}
+ \frac{\partial f_i(x,t)}{\partial t} \Delta t  \nonumber \\
+ \frac{\partial^2
  f_i(x,t)}{\partial t^2} \frac{\Delta t^2}{2} + \frac{\partial^2
  f_i(x,t)}{\partial x\partial t} i\Delta x \Delta t + \ldots. 
\end{eqnarray}
Combined
with the assumption that $f_i$ is a solution of the LBE
\eqref{LBE} on an infinite domain, we end up with expansion
\begin{eqnarray}
f_i(x,t)=f_i^{eq}(x,t)-\frac{i\Delta x}{\omega}\frac{\partial f_i(x,t)}{\partial x}-\frac{i^2 \Delta x^2}{2\omega}\frac{\partial^2 f_i(x,t)}{\partial x^2}-\frac{\Delta t}{\omega}\frac{\partial f_i(x,t)}{\partial t} \nonumber \\
-\frac{\Delta t^2}{2\omega}\frac{\partial^2 f_i(x,t)}{\partial t^2}-\frac{i\Delta x \Delta t}{\omega} \frac{\partial^2 f_i(x,t)}{\partial x \partial t}-\ldots,
\end{eqnarray}
and higher order terms. With the notation $\mathcal{L}_i$ for the functional
\begin{equation} \label{functional}
\mathcal{L}_i:=-\frac{i\Delta x}{\omega}\frac{\partial }{\partial x}-\frac{i^2 \Delta x^2}{2\omega}\frac{\partial^2 }{\partial x^2}-\frac{\Delta t}{\omega}\frac{\partial }{\partial t}-\frac{\Delta t^2}{2\omega}\frac{\partial^2 }{\partial t^2} -\frac{i\Delta x \Delta t}{\omega} \frac{\partial^2 }{\partial x \partial t}-\ldots,
\end{equation}
we can rewrite the LBE into a set of three coupled PDEs for the distribution functions
\begin{equation}
(1-\mathcal{L}_i) f_i(x,t)=f_i^{eq}(x,t), \quad  \forall i \in \{-1,0,1\},
\end{equation} 
that holds for $x \in
]-\infty,\infty[$ and $t \in [0,\infty[$.

The solution can be found
by performing a Picard or fixed point iteration 
\begin{equation}\label{eq:picarditeration}
f_i^{(n+1)}(x,t) = \mathcal{L}_if_i^{(n)}(x,t) + f_i^{eq}(x,t),
\end{equation}
with initial guess $f_i^{(-1)}(x,t) = 0$. The solution is then 
$f_i(x,t) = \sum_{k=0}^\infty \mathcal{L}_i^k f_i^{eq}(x,t)$. This series converges if the
error between subsequent iterations goes to zero. This can be
satisfied when smoothness on $f_i^{eq}$ is guaranteed such that $\lim_{n
  \rightarrow \infty} \| \mathcal{L}_i^{n+1} f_i^{eq}\| = 0$
\cite{vanderhoydonc2}.

This derivation shows that Eq.~\eqref{NCHE_rho_rhou} is valid when the derivatives of $f_i^{eq}$ are easily expressible as derivatives of the conserved moments.

If a PDE in closed form exists that describes the time derivatives of
  $\rho$ and $\rho u$ in only spatial derivatives of $\rho$ and $\rho u$, then it is possible to eliminate $\partial_t \rho$ and $\partial_t (\rho u)$ from the expansion. The distribution functions can then be approximated by a truncated series with only spatial derivatives and becomes 
\begin{equation} \label{exp_spatial_rho_rhou}
  f_i(x,t)  =  f_i^{eq}(x,t) + \alpha_i^{(1)} \partial_x \rho + \alpha_i^{(2)} \partial_x (\rho u) +  \beta_i^{(1)}  \partial_x^2 \rho  + \beta_i^{(2)} \partial_x^2 (\rho u).
\end{equation}

The coefficients $\alpha_i^{(1)}$, $\alpha_i^{(2)}$, $\beta_i^{(1)}$ and $\beta_i^{(2)}$ can then be extracted from a linear system analogous to the one in Eq.~\eqref{system_constants_spatial} (extended for the multiple conserved moments).
The algorithm to obtain the coefficients from this linear system is similar to the one in the previous section. 
% The Constrained Runs algorithm is combined with Newton's method to find the distribution functions closer to the slow manifold. 

The construction of the numerical Chapman--Enskog expansion is based on an expansion that expresses the distribution functions as a series of the macroscopic moments. The CR algorithm is the underlying method to determine the coefficients of this series. In order to generalize the numerical Chapman--Enskog expansion for density and momentum we also need to generalize the CR algorithm. This is outlined in Section \ref{CR_gen}.

\subsection{Constrained Runs algorithm} \label{CR_gen}
The original CR algorithm is performed on the velocity moments. The intention of this generalization is to perform the algorithm directly on the distribution functions. Initially, we start from distribution functions $(f_{1}^0,f_0^0,f_{-1}^0)^T$. The CR algorithm is based on a backward extrapolation in time of the missing moments. For this, we need to evolve the distribution functions over some time steps. This results in $(f_{1}^1,f_0^1,f_{-1}^1)^T$, $(f_{1}^2,f_0^2,f_{-1}^2)^T$, $\ldots$.
$\boldsymbol{M}$ and $\boldsymbol{M}^{-1}$ are applied to convert distribution functions to moments and vice versa. Then, the CR algorithm resets the conserved moments, namely density and momentum, and applies the smoothness condition on the remaining moment. 
To reset the conserved moments density and momentum and apply the smoothness condition on the distribution functions, we use
% \begin{equation} \label{gen_eq_CR}
% \left(\begin{array}{c} f_1 \\
% f_0 \\
% f_{-1} \end{array} \right)^{next}=\left(\begin{array}{c} f_1 \\
% f_0 \\
% f_{-1} \end{array} \right)^{prev}+\boldsymbol{M}^{-1}\boldsymbol{M}^0 \left(\left(\begin{array}{c} f_1 \\
% f_0 \\
% f_{-1} \end{array} \right)^0-\left(\begin{array}{c} f_1 \\
% f_0 \\
% f_{-1} \end{array} \right)^{prev} \right),
% \end{equation}
% where
% \begin{equation*}
% \boldsymbol{M}^0 = \left(\begin{array}{c c c}
% 1 & 1 & 1 \\
% 1 & 0 & -1 \\
% 0 & 0 & 0 \end{array} \right),
% \end{equation*}
\begin{equation} \label{gen_eq_CR}
\left(\begin{array}{c} f_1 \\
f_0 \\
f_{-1} \end{array} \right)^{next}=\left(\begin{array}{c} f_1 \\
f_0 \\
f_{-1} \end{array} \right)^{prev}+\boldsymbol{M}^{-1}\boldsymbol{M}^0 \left(\left(\begin{array}{c} f_1 \\
f_0 \\
f_{-1} \end{array} \right)^0-\left(\begin{array}{c} f_1 \\
f_0 \\
f_{-1} \end{array} \right)^{prev} \right),
\end{equation}
where
\begin{equation*}
\boldsymbol{M}^0 = \left(\begin{array}{c c c}
1 & 1 & 1 \\
1 & 0 & -1 \\
0 & 0 & 0 \end{array} \right),
\end{equation*}
and $(f_{1}^{prev},f_0^{prev},f_{-1}^{prev})^T$ contains the backward extrapolation in time applied on the distribution functions.
For example, for a linear extrapolation in time $(f_{1}^{prev},f_0^{prev},f_{-1}^{prev})^T=2(f_{1}^{1},f_0^{1},f_{-1}^{1})^T - (f_{1}^{2},f_0^{2},f_{-1}^{2})^T $. The distribution functions are then eventually determined via an iterative method performed on Eq.~\eqref{gen_eq_CR}.

\ \\
Here, we return to the derivation of the numerical Chapman--Enskog expansion. We need to determine $\boldsymbol{\alpha}^{(1)}$, $\boldsymbol{\alpha}^{(2)}$, $\boldsymbol{\beta}^{(1)}$ and $\boldsymbol{\beta}^{(2)}$ such that they are invariant under the numerical function $h(\boldsymbol{\alpha}^{(1)},\boldsymbol{\alpha}^{(2)},\boldsymbol{\beta}^{(1)},\boldsymbol{\beta}^{(2)}; \rho, \rho u,m)$ given in Function \ref{fun:h2}.

\begin{algorithm}[h]
\floatname{algorithm}{Function}
  \caption{$h(\boldsymbol{\alpha}^{(1)},\boldsymbol{\alpha}^{(2)},\boldsymbol{\beta}^{(1)},\boldsymbol{\beta}^{(2)}; \rho, \rho u,m)$ \label{fun:h2}} 
  \label{algo:coefficients2}
  \begin{algorithmic}[1]
  \REQUIRE Guess on vectors of coefficients $\boldsymbol{\alpha}^{(1)}$, $\boldsymbol{\alpha}^{(2)}$, $\boldsymbol{\beta}^{(1)}$, $\boldsymbol{\beta}^{(2)}$, given density $\rho$ and momentum $\rho u$, order of smoothness condition $m$.
    \STATE Construct lifting operator in Eq.~\eqref{exp_spatial_rho_rhou}.
    \STATE Apply Eq.~\eqref{gen_eq_CR}. This results in new distribution functions closer to the slow manifold.
    \STATE Select grid points to construct a similar linear system as in \eqref{system_constants_spatial}.
    \STATE Solve the system for $\boldsymbol{\alpha}^{(1)}$, $\boldsymbol{\alpha}^{(2)}$, $\boldsymbol{\beta}^{(1)}$ and $\boldsymbol{\beta}^{(2)}$.
    \RETURN $\boldsymbol{\alpha}^{(1)}$, $\boldsymbol{\alpha}^{(2)}$, $\boldsymbol{\beta}^{(1)}$, $\boldsymbol{\beta}^{(2)}$.
  \end{algorithmic}
\end{algorithm}

Function \ref{algo:coefficients} and \ref{algo:coefficients2} mostly correspond to each other. Although, Function \ref{algo:coefficients2} is more general. It contains multiple conserved moments, the extended numerical Chapman--Enskog expansion and a generalized step of the CR algorithm.

This section shows that the numerical Chapman--Enskog expansion can be generalized in this setting. However, for this to be possible, the derivatives of the equilibrium distribution should be easily expressible as derivatives of the conserved moments.

\section{Numerical results} \label{numerical_results}

\subsection{Example 1: D1Q3}
As a model problem, we consider a D1Q3 lattice Boltzmann model with equilibrium distribution
 \begin{equation*}
 f_1^{eq}(x,t) = \left(\frac{1}{3}+\frac{1}{2c} u(x,t)\right) \rho(x,t), \quad f_0^{eq}(x,t) = \frac{1}{3}\rho(x,t), \quad
f_{-1}^{eq}(x,t) =  \left(\frac{1}{3}-\frac{1}{2c} u(x,t)\right) \rho(x,t),
\end{equation*}
where $c=\Delta x / \Delta t$.
Note that the equilibrium depends on the density and momentum as opposed to only the density in \cite{vanderhoydonc2}.
This equilibrium is built such that the zeroth and first order velocity moments
 are conserved. It just represents a simple model problem to test the numerical Chapman--Enskog expansion on such an extended problem.

The considered model problem has the following parameters for a one-dimensional domain of length $L$ with $n$ spatial grid points.
\begin{eqnarray} \label{model_problem}
& & L = 10, \quad n = 200, \quad \Delta x = L/n = 0.05, \quad  \Delta t = 0.001, \quad \omega=0.9091, \nonumber \\
& & \rho(x,0) = \exp(-(x-L/2)^2) + 0.1, \quad u(x,0) = 0.03\sin (2\pi x/L).
\end{eqnarray}
Furthermore, periodic boundary conditions are taken into account.

The lifting operator can be tested against a reference distribution
function $f_c$.  This reference solution is calculated by performing
1000 lattice Boltzmann steps starting from an initial state that
corresponds to the equilibrium distribution function with given
macroscopic variables $\rho(x,0)$ and $u(x,0)$.

The lifting operator can be evaluated by restricting the reference
distribution function $f_c$ to its macroscopic variables and lift them
back to a distribution function $f$ by using the numerical
Chapman--Enskog expansion. The resulting  $f$ is compared with
$f_{c}$ with the help of the 2-norm $\|f-f_{c}\|$. These results are
included in Table \ref{table_restrict_lift} for various orders of the
expansion and different orders of the underlying CR algorithm. As a
comparison, lifting with the corresponding equilibrium distribution has a 2-norm of $\|f^{eq}-f_{c}\|=0.0415$.

We calculate the lifting operator as a fixed point for the coefficients and thus perform a Newton iteration on the coefficients of the numerical Chapman--Enskog expansion by using the underlying CR algorithm to find the distribution functions closer to the slow manifold.
As in \cite{vanderhoydonc2}, we know that the coefficients can be determined for every density and momentum as long as the problem has similar parameters $\Delta x$ and $\Delta t$. Thus, we choose macroscopic variables such that the linear system from which the coefficients are determined is not singular.

\begin{table}[h]
\caption{Test lifting operator against a reference distribution function $f_c$. The error $\|f-f_c\|$ is presented in which $f$ is determined via the numerical Chapman--Enskog expansion. The reference distribution function $f_c$ is obtained by performing 1000 LBM time steps on the initial state with parameters listed in \eqref{model_problem}. These results are listed for increasing number of terms in the expansion, implying an increasing number of considered coefficients. The order of the underlying CR algorithm also increases for every considered expansion. \label{table_restrict_lift}} 
\begin{center} \footnotesize
\begin{tabular*}{\hsize}{@{\extracolsep{\fill}}llllllll@{}}
\hline
% $f^{eq}$, $\partial_x \rho$ &  0  & 0.0415 \\
% & 1 & 0.0417 \\
% & 2 & 0.0415 \\
% & 3 & 0.0415 \\
% & 4 & 0.0415 \\
% & 5 & 0.0415 \\
% & 6 & 0.0415 \\ \hline
Order CR &  0 & 1 & 2 & 3 & 4 & 5 & 6 \\ \hline
\multicolumn{8}{l}{ $f^{eq}$, $\partial_x \rho$, $\partial_x (\rho u)$  } \\ 
$\|f-f_c\|$  & 0.0679 & 0.0320 &  0.0071 & 0.0074 & 0.0073 & 0.0073 & 0.0073 \\ \hline 
\multicolumn{8}{l}{ $f^{eq}$, $\partial_x \rho$, $\partial_x (\rho u)$, $\partial_x^2 \rho$, $\partial_x^2 (\rho u)$  } \\ 
$\|f-f_c\|$  &  0.0861  & 0.0026 & 2.6378e-004 & 6.7372e-005 & 3.7216e-005 & 3.6750e-005 & 3.6798e-005 \\ \hline 
\multicolumn{8}{l}{$f^{eq}$, $\partial_x \rho$, $\partial_x (\rho u)$, $\partial_x^2 \rho$, $\partial_x^2 (\rho u)$, $\partial_x^3 \rho$, $\partial_x^3 (\rho u)$ } \\ 
$\|f-f_c\|$  & 0.0861  & 0.0027 & 5.6094e-004 & 8.2239e-005 & 4.2439e-006 & 5.0278e-006 & 5.3527e-006 \\ \hline 
\multicolumn{8}{l}{$f^{eq}$, $\partial_x \rho$, $\partial_x (\rho u)$, $\partial_x^2 \rho$, $\partial_x^2 (\rho u)$, $\partial_x^3 \rho$, $\partial_x^3 (\rho u)$, $\partial_x^4 \rho$, $\partial_x^4 (\rho u)$ } \\ 
$\|f-f_c\|$  & 0.0861  & 0.0027 & 3.8770e-004 & 1.6050e-005 & 2.1631e-006 & 2.6841e-007 & 6.5869e-008 \\ \hline
\end{tabular*}
\end{center}
\end{table}

Note that taking more terms into account in the expansion increases the accuracy but brings down the beauty of this method since the number of coefficients increases significantly. 

As a comparison we show the CR results for this model problem. For the results in Table \ref{CR_example1} we apply CR on the distribution functions. In order to avoid instabilities, the CR algorithm is combined with Newton's method. This results in a Jacobian of size $600 \times 600$. For this small test problem, it is still possible to apply the original CR algorithm so it is nice to compare these results with those obtained via the numerical Chapman--Enskog expansion. However --- as the size of the Jacobian clarifies --- the CR algorithm can not be applied for higher dimensional problems.

\begin{table}[h]
\caption{Test Constrained Runs algorithm combined with Newton's method directly applied on distribution functions against a reference distribution $f_c$. The error $\|f-f_c\|$ is presented in which $f$ is the lifting operator obtained via the CR algorithm. The reference distribution function $f_c$ is obtained by performing 1000 LBM time steps on the initial state with parameters listed in \eqref{model_problem}. The results are presented for different orders of the smoothness condition for the CR algorithm. \label{CR_example1}}
\begin{center} \footnotesize
\begin{tabular*}{\hsize}{@{\extracolsep{\fill}}llllllll@{}}
\hline
Order CR & 0 & 1 & 2 & 3 & 4 & 5 & 6 \\ \hline 
$\|f-f_c\|$ &  0.0861 &  0.0027 & 3.8573e-004 &  1.9192e-005 &  2.5547e-006  &  2.8042e-007 &  2.1958e-008 \\
\hline
\end{tabular*}
\end{center}
\end{table}

\subsection{Example 2: D2Q5}

The model problem for this example contains five velocity directions in a two-dimensional spatial domain. These are given by $v_0=(0,0)$, $v_1=(+c,0)$, $v_2=(0,+c)$, $v_3=(-c,0)$ and $v_4=(0,-c)$ with $c=\Delta x/\Delta t=\Delta y/\Delta t$. The following equilibrium distribution conserves density and average flow velocities in the $x$ and $y$ direction.
\begin{eqnarray*}
f^{eq}(x,y,v_0,t) &=& \frac{1}{5}\rho(x,y,t), \quad
f^{eq}(x,y,v_1,t) = \left(\frac{1}{5} + \frac{1}{2c}u_x(x,y,t) \right) \rho(x,y,t), \\ \nonumber
f^{eq}(x,y,v_2,t) &=& \left(\frac{1}{5} + \frac{1}{2c}u_y(x,y,t) \right) \rho(x,y,t), \quad
f^{eq}(x,y,v_3,t) = \left(\frac{1}{5} - \frac{1}{2c}u_x(x,y,t) \right) \rho(x,y,t), \\ \nonumber
f^{eq}(x,y,v_4,t) &=& \left(\frac{1}{5} - \frac{1}{2c}u_y(x,y,t) \right) \rho(x,y,t).
\end{eqnarray*}

The considered model problem has the following parameters for a two-dimensional domain of size $L \times L$ with $n^2$ spatial grid points.
\begin{eqnarray} \label{model_problem2}
& & L = 10, \quad n = 200, \quad \Delta x = L/n = 0.05, \quad \Delta y = L/n = 0.05, \quad  \Delta t = 0.001, \quad \omega=0.9091, \nonumber \\
& & \rho(x,y,0) = \exp(-(x-L/2)^2-(y-L/2)^2)+0.4, \quad u_x(x,y,0) = 0.03\sin (2\pi x/L), \nonumber \\ 
& & u_y(x,y,0) = 0.03\sin (2\pi y/L).
\end{eqnarray}
Furthermore, periodic boundary conditions are taken into account.

Again we test the numerical Chapman--Enskog expansion in a setting of restriction and lifting. A reference distribution function $f_c(x,y,v_i,t)$, $i \in \{0,1,2,3,4\}$ is obtained by performing 1000 LBM time steps on the initial state given in Eq.~\eqref{model_problem2}. A possible lifting strategy that we can compare with is lifting with the corresponding equilibrium distribution. The 2-norm errors  for the specific velocities are listed below. 
\begin{eqnarray*}
\|f^{eq}(x,y,v_0,t)-f_{c}(x,y,v_0,t)\| &=& 0.0470, \\
\|f^{eq}(x,y,v_i,t)-f_{c}(x,y,v_i,t)\| &=& 0.0550, \quad i \in \{1,\ldots,4\}.
\end{eqnarray*}
% \begin{equation*}
% \|f^{eq}(x,y,v_0,t)-f_{c}(x,y,v_0,t)\| = 0.0470, \quad \|f^{eq}(x,y,v_i,t)-f_{c}(x,y,v_i,t)\| = 0.0550, \quad i \in \{1,\ldots,4\}.
% \end{equation*}
Lifting with the numerical Chapman--Enskog expansion is listed in Table \ref{test_lifting_D2Q5}.

\begin{table}[h]
\caption{Test lifting operator against a reference distribution function $f_c$. The errors $\|f(x,y,v_i,t)-f_c(x,y,v_i,t)\|$ for $i\in \{0,\ldots,4\}$ are presented in which $f$ is determined via the numerical Chapman--Enskog expansion. The reference distribution function $f_c$ is obtained by performing 1000 LBM time steps on the initial state with parameters listed in \eqref{model_problem2}. These results are listed for increasing number of terms in the expansion, implying an increasing number of considered coefficients. The order of the underlying CR algorithm also increases for every considered expansion. \label{test_lifting_D2Q5}} 
\begin{center} \footnotesize
\begin{tabular*}{\hsize}{@{\extracolsep{\fill}}llllllll@{}} \hline
% \multicolumn{4}{@{\extracolsep{\fill}}l}{$f^{eq}$, $\partial_x \rho$, $\partial_y \rho$, $\partial_x (\rho u_x)$, $\partial_y (\rho u_x)$, $\partial_x (\rho u_y)$, $\partial_y (\rho u_y)$} \\ \hline
Order CR & 0  & 1  &  2 &  3 & 4  & 5  &  6  \\ \hline
\multicolumn{8}{l}{$f^{eq}$, $\partial_x \rho$, $\partial_y \rho$, $\partial_x (\rho u_x)$, $\partial_y (\rho u_x)$, $\partial_x (\rho u_y)$, $\partial_y (\rho u_y)$} \\ \hline
$\|f(x,y,v_0,t)-f_{c}(x,y,v_0,t)\|$ & 0.6618 & 0.9794 & 0.1189 &  0.1179 & 0.1179 & 0.1179 & 0.1179  \\
$\|f(x,y,v_1,t)-f_{c}(x,y,v_1,t)\|$ & 0.1530 & 0.2294 & 0.0143 &   0.0140 & 0.0140 & 0.0140 & 0.0140  \\
$\|f(x,y,v_2,t)-f_{c}(x,y,v_2,t)\|$ &0.1779 & 0.2603 & 0.0452 &   0.0450 & 0.0450 & 0.0450 & 0.0450   \\
$\|f(x,y,v_3,t)-f_{c}(x,y,v_3,t)\|$ &0.1530 & 0.2294 & 0.0143 &   0.0140 & 0.0140 & 0.0140 & 0.0140   \\
$\|f(x,y,v_4,t)-f_{c}(x,y,v_4,t)\|$ &0.1779 & 0.2603 & 0.0452 &   0.0450 & 0.0450 &  0.0450 & 0.0450    \\ \hline \hline
% \multicolumn{4}{@{\extracolsep{\fill}}l}{$f^{eq}$, $\partial_x \rho$, $\partial_y \rho$, $\partial_x (\rho u_x)$, $\partial_y (\rho u_x)$, $\partial_x (\rho u_y)$, $\partial_y (\rho u_y)$, $\partial_x^2 \rho$, $\partial_y^2 \rho$, $\partial_x^2 (\rho u_x)$, $\partial_y^2 (\rho u_x)$, $\partial_x^2 (\rho u_y)$, $\partial_y^2 (\rho u_y)$} \\ \hline
\multicolumn{8}{l}{$f^{eq}$, $\partial_x \rho$, $\partial_y \rho$, $\partial_x (\rho u_x)$, $\partial_y (\rho u_x)$, $\partial_x (\rho u_y)$, $\partial_y (\rho u_y)$, $\partial_x^2 \rho$, $\partial_y^2 \rho$, $\partial_x^2 (\rho u_x)$, $\partial_y^2 (\rho u_x)$, $\partial_x^2 (\rho u_y)$, $\partial_y^2 (\rho u_y)$} \\ \hline
$\|f(x,y,v_0,t)-f_{c}(x,y,v_0,t)\|$ & 0.1928 & 0.0176 & 5.2067e-004 &  9.7954e-005 & 9.9099e-005 & 9.8667e-005 & 9.8663e-005 \\
$\|f(x,y,v_1,t)-f_{c}(x,y,v_1,t)\|$ & 0.0483 & 0.0044 & 1.2264e-004 &  6.2397e-005  & 6.2491e-005 & 6.2455e-005 & 6.2452e-005   \\
$\|f(x,y,v_2,t)-f_{c}(x,y,v_2,t)\|$ & 0.0483 & 0.0044 & 1.4786e-004 &  5.7104e-005 & 5.7140e-005 & 5.7129e-005 & 5.7129e-005   \\
$\|f(x,y,v_3,t)-f_{c}(x,y,v_3,t)\|$ & 0.0483 & 0.0044 & 1.2264e-004 &  6.2397e-005  & 6.2491e-005 & 6.2455e-005 & 6.2452e-005   \\
$\|f(x,y,v_4,t)-f_{c}(x,y,v_4,t)\|$ & 0.0483 & 0.0044 & 1.4786e-004 &  5.7104e-005  & 5.7140e-005 & 5.7129e-005 & 5.7129e-005   \\ \hline \hline
% \multicolumn{4}{@{\extracolsep{\fill}}l}{$f^{eq}$, $\partial_x \rho$, $\partial_y \rho$, $\partial_x (\rho u_x)$, $\partial_y (\rho u_x)$, $\partial_x (\rho u_y)$, $\partial_y (\rho u_y)$, $\partial_x^2 \rho$, $\partial_y^2 \rho$, $\partial_x^2 (\rho u_x)$, $\partial_y^2 (\rho u_x)$, $\partial_x^2 (\rho u_y)$, $\partial_y^2 (\rho u_y)$, $\partial_x^3 \rho$, $\partial_y^3 \rho$, $\partial_x^3 (\rho u_x)$, $\partial_y^3 (\rho u_x)$, $\partial_x^3 (\rho u_y)$, $\partial_y^3 (\rho u_y)$} \\ \hline
\multicolumn{8}{l}{$f^{eq}$, $\partial_x \rho$, $\partial_y \rho$, $\partial_x (\rho u_x)$, $\partial_y (\rho u_x)$, $\partial_x (\rho u_y)$, $\partial_y (\rho u_y)$, $\partial_x^2 \rho$, $\partial_y^2 \rho$, $\partial_x^2 (\rho u_x)$, $\partial_y^2 (\rho u_x)$, $\partial_x^2 (\rho u_y)$, $\partial_y^2 (\rho u_y)$, $\partial_x^3 \rho$, $\partial_y^3 \rho$, }  \\ 
\multicolumn{8}{l}{$\partial_x^3 (\rho u_x)$, $\partial_y^3 (\rho u_x)$, $\partial_x^3 (\rho u_y)$, $\partial_y^3 (\rho u_y)$} \\\hline
$\|f(x,y,v_0,t)-f_{c}(x,y,v_0,t)\|$ & 0.1928 & 0.0178 & 6.7160e-004 &  2.1868e-004 & 5.4479e-005 & 5.4150e-005 & 5.4292e-005   \\
$\|f(x,y,v_1,t)-f_{c}(x,y,v_1,t)\|$ & 0.0483 & 0.0045 & 1.4442e-004 &  7.0300e-005  & 1.7193e-005 & 1.7206e-005 &  1.7215e-005   \\
$\|f(x,y,v_2,t)-f_{c}(x,y,v_2,t)\|$ & 0.0483 & 0.0045 & 1.9339e-004 &  4.1264e-005 & 2.6411e-005 & 2.6403e-005 & 2.6416e-005   \\
 $\|f(x,y,v_3,t)-f_{c}(x,y,v_3,t)\|$ & 0.0483 & 0.0045 & 1.4442e-004 &  7.0300e-005 & 1.7193e-005 & 1.7206e-005 & 1.7215e-005   \\
$\|f(x,y,v_4,t)-f_{c}(x,y,v_4,t)\|$ & 0.0483 & 0.0045 & 1.9339e-004 &  4.1264e-005 & 2.6411e-005 & 2.6403e-005 & 2.6416e-005   \\ \hline 
\end{tabular*}
\end{center}
\end{table}
Again, we see an increase in the accuracy when more terms of the expansion are taken into account. However, the increase in terms of the expansion brings down the beauty of this method. Especially since all the cross terms between the different spatial directions should also be considered to obtain a better accuracy.

\section{Conclusions} \label{conclusions}
This paper constructs a numerical Chapman--Enskog expansion that can be
used as a lifting operator for LBMs that conserve density and momentum
during the collision. This is an extension of \cite{vanderhoydonc2}
where the idea of a numerical Chapman--Enskog expansion was first
introduced and applied to LBMs that only conserve the density.  This method is an alternative to
the Chapman--Enskog expansion that requires a tedious analytical derivation. 

The new method was shown to converge for some simple benchmark
problems in 1D and 2D. The numerical results show that the method is
computationally much cheaper than the Constrained Runs algorithm, an
alternative numerical lifting operator for LBMs. The latter becomes so expensive that it is too prohibitive
to apply in 2D. The extension of the numerical Chapman--Enskog expansion proposed in
this paper required a generalization of the CR algorithm that forms the
basis of the method. Unfortunately, it is necessary to assume that the derivatives of the equilibrium distribution of the LBM are easily expressible as derivatives of the conserved moments.

In addition, we saw a significant increase in the number of coefficients
that need to be determined when the numerical Chapman--Enskog expansion
includes higher order terms.  Similarly, an expansion in 2D requires many additional
cross terms that appear when these higher order terms are used in the
expansion. Each additional term introduces new unknown coefficients
that need to be determined by Newton's method. However, the resulting system
is still much smaller than the system constructed in the
original CR algorithm.

In future research we will try to remove the limiting
assumptions and generalize the expansion to kinetic models that use Maxwell--Boltzmann
equilibrium distributions that are frequently used in physical applications.

\section*{Acknowledgements}

This work is supported by research project \textit{Hybrid macroscopic
  and microscopic modelling of laser evaporation and expansion},
G.017008N, funded by `Fonds Wetenschappelijk Onderzoek' together with
an `ID-beurs' of the University of Antwerp.

%% References


\bibliographystyle{elsarticle-num} % NUMBERED IN ORDER OF APPEARANCE
\bibliography{biblio_procedia}






% \end{thebibliography}



% \appendix
% \section{An example appendix}
% Authors including an appendix section should do so after References section. Multiple appendices should all have headings in the style used above. They will automatically be ordered A, B, C etc.

% \subsection{Example of a sub-heading within an appendix}
% There is also the option to include a subheading within the Appendix if you wish.

\end{document}